Preparing to Integrate Generative Pretrained Transformer Series 4 models into Genetic Variant Assessment Clinical Workflow: Assessing Performance, Drift, and Nondeterminism Characteristics Relative to Classifying Functional Evidence in Literature

Samuel J. Aronson[1,2,*,±], Kalotina Machini[1,3,*], Jiyeon Shin[1,2*], Pranav Sriraman[1], Sean Hamill[4], Emma R. Henricks[1], Charlotte J. Mailly[1,2], Angie J. Nottage[1], Sami S. Amr[1,3], Michael Oates[1,2], Matthew S. Lebo[1,3]

1 Mass General Brigham Personalized Medicine
2 Accelerator for Clinical Transformation
3 Department of Pathology, Brigham and Women's Hospital
4 Microsoft Corporation
* These authors contributed equally
± Corresponding author

## Abstract

Background. Large Language Models (LLMs) hold promise for improving genetic variant literature review in clinical genetic testing. We assessed Generative Pretrained Transformer 4 (GPT-4) series models' performance, nondeterminism, and drift to inform decisions regarding their present suitability for use in complex clinical laboratory processes.

Methods. A two-prompt chain of thought GPT-4 based process for automated classification of functional evidence in literature was optimized using a development set of 45 articles paired with genetic variants. The prompts first asked GPT-4 to supply all functional evidence present in an article related to the variant of interest or indicate that no such functional evidence is present. For articles that GPT-4 indicated contained functional evidence, a second prompt asked GPT-4 to classify the evidence into pathogenic, benign, or intermediate/inconclusive categories. A final test set of 72 manually classified articles paired with genetic variants was used to test performance over time.

Results. Over a 2.5-month period from December 2023-February 2024, we observed substantial differences in intraday results (nondeterminism) and across days (drift) which lessened after 1/18/24. This variability existed both within and across models in the GPT-4 series, affecting different performance statistics (e.g., sensitivity, positive predictive value [PPV], and negative predictive value [NPV]) to different degrees. Twenty runs of our test set after 1/18/24 identified articles with functional evidence with 92.2% sensitivity, 95.6% PPV and 86.3% NPV. Our second prompt's pathogenic evidence detection had 90.0% sensitivity, 74.0% PPV and 95.3% NVP and its benign evidence detection had 88.0% sensitivity, 76.6% PPV and 96.9% NVP.

Conclusion. Nondeterminism within LLMs must be assessed and drift should be monitored for the specific metrics used to decide whether to introduce LLM based functionality into clinical workflows. Failing to do this assessment or accounting for these challenges could lead to incorrect or missing information that is critical for patient care. The performance of our prompts appears adequate to assist in article prioritization but not in automated decision making. Multiple avenues for further enhancing this performance could be explored.

## Key Words
Variant classification; artificial intelligence; GPT-4; large language models; functional evidence; Nondeterminism; model drift

Introduction

Clinical interpretation of genetic variants identified in patients is a multi-step process relying on accurate integration of diverse types of evidence. Some of the strongest evidence for or against pathogenicity, according to current guidelines set by the American College of Medical Genetics (ACMG) and the Association for Molecular Pathology (AMP), include genetic and functional data. These data elements are primarily collected through literature search and review, which can be a very time-consuming process typically conducted by highly specialized and expensive staff. Functional studies, specifically, assess the effects of a variant on protein function such as localization, enzymatic activity, and electrophysical capacity using both *in vivo* and *in vitro* models[1–3]. Further, current technologies allow for high-throughput screening of functional assays, greatly expanding the number of variants with validated functional evidence[4,5]. Because functional data identified through literature review is critical for variant classification, it is essential that such evidence is robustly and accurately collected and interpreted. Assisting personnel in their review of functional evidence from literature can decrease the time, and therefore the cost, of this process, thereby improving scalability.

Large Language Models (LLMs), exemplified by technologies such as Generative Pre-trained Transformer version 4 (GPT-4)[6], are a recent development in artificial intelligence (AI) that have shown high performance in their ability to summarize written text. However, they are subject to hallucinations – seemingly factual statements without a validated source document[7] – and at times provide inaccurate information. Deploying them in clinical workflows involves many different components, some of which may change over time. This can lead to "drift" in model performance. GPT-4 is also currently nondeterministic, with repeated runs under what appear to be the same conditions producing different results (Figure 1)[8].

Therefore, while LLM are potentially powerful tools, their accuracy and temporal performance characteristics must be validated and vetted, especially before use in a clinical laboratory process. If these characteristics can be optimized, they could assist the variant assessment workflow either by prioritizing or deprioritizing articles for manual review so critical information is found faster. If specific metrics reach extremely high levels, it may become possible to rely on automated assessment in certain circumstances. For example, if NPV for assessments of functional data or genetic evidence approach 100%, articles could potentially be excludable from manual review as opposed to simply deprioritized.

Here, we assess whether a chain-of-thought GPT-4 prompting strategy can identify if an article contains functional data relative to a specific variant, classify the nature of functional data that is detected, and return an output in a manner that could be machine readable through regular expression matching, as well as monitoring the accuracy of this process over multiple runs and days.

Methods

A training dataset consisting of 45 articles was developed, each evaluated for a single variant. For each of the 45 variant-paper pairs, a geneticist and/or a genetic counselor manually assessed whether functional evidence was present in the article for the specific variant, and if so, whether the functional evidence indicated support for or against pathogenicity using the following categories: "Pathogenic Evidence," "Evidence of Intermediate Function", "Assays are Inconclusive" or "Benign Evidence". The training set consisted of the article, the variant assessed, and the manually curated conclusion. Article-variant pairs were obtained from Human Gene Mutation Database (HGMD) Professional (v2023.3) and the internal knowledgebase used for managing genomic variant classification at Mass General Brigham Personalized Medicine (MGBPM).

Initially, we experimented with prompt sequences using Bing Chat's version of GPT-4. We switched to using Azure AI GPT-4 Application Programming Interfaces (APIs) when they became available to us. We developed a python program to execute prompts with the temperature parameter, which specifies a level of randomness to interject into responses, set to 0.

We iteratively refined and tested our prompt sequences against the training dataset. We started by using longer prompt sequences to "walk" GPT-4 through the required logic. We then focused on reducing the number of prompts required to achieve equivalent results, arriving at a system prompt and a two-step prompt sequence (Supplemental Table 1 and Figure 2). This sequence first determines if functional evidence is present in the publication for the specified variant and then classifies the functional evidence for the variant if determined to be present. Because variant nomenclatures are inconsistent in the literature, the first prompt was run up to three times using different variant nomenclature to attempt to capture different representations for the functional data (Supplement Figure 1).

After each attempt, the system uses a regular expression to check whether GPT's response indicates that functional evidence is present. If functional evidence has not been identified after the different nomenclature prompts, "Assays Not Present" is returned. If functional evidence is identified, the second prompt is run and a post processing script takes the final outputs from GPT and uses regular expression matching to classify the result and output according to the following 4 categories: "Pathogenic Evidence," "Evidence of Intermediate Function", "Assays are Inconclusive" or "Benign Evidence". In our analysis we labeled articles where our regular expressions failed to match GPT's output as "Cannot Classify". In our analysis "Cannot Classify" and "Assays Not Present" where merged to create the "No Assays Detected" category.

The python program's input consisted of: (1) a pdf file representation of the article, (2) the target gene symbol, (3) the target variant in c.<nucleotide change> (p.<Amino Acid change>) format, for example: c.3694C>T (p.R1232W), and (4) a list of nomenclature aliases equivalent to the target variant. We obtained nomenclature aliases by passing the genome build, gene transcript, and nucleotide change to the VariantValidator API[9]. These aliases included genomic changes formatted in build GRCh37 and GRCh38 and HGVS standard format. In addition, we incorporated short and long amino acid forms (e.g., p.Arg1232Trp and p.R1232W), with and without "p." and parenthesis and with and without "c." for the nucleotide change. The PDF versions of the articles were converted to text using PyPDF2 v3.0.1 and the resulting text was embedded in the prompt sequence. Of note, due to challenges with token size and GPT-4 model use, the pipeline was not designed to include information from figures and supplemental materials, though figure legends and manuscript tables were included.

An independent final test set of 72 article-variant pairs was created. Each variant-article pair was labeled by a geneticist and/or a genetic counselor using these categories: "Pathogenic Evidence," "Evidence of Intermediate Function," "Assays are Inconclusive", "Benign Evidence," or "Assays Not Present". Because of the ambiguity manually differentiating between intermediate and inclusive evidence, we group these two responses together into a combined category in our results. The articles in the final test set were not used for training purposes. Our program ran in a HIPAA compliant environment that did not allow Microsoft of OpenAI to persist our prompts and responses for future training. As our pipeline does not process information in figures and supplemental materials, articles where variant-level functional evidence was only included in these formats were not included in our test set. For the many articles in our test set with information in both the main text/tables and in figures/supplemental materials, only the information in the main text and tables was assessed, potentially disadvantaging GPT-4 performance. Supplemental Table 2 contains our test dataset, its labels, and the results of reach GPT-4 series run. The table is normalized to one-word results as opposed to more verbose categories.

We ran the test set repeatedly over time ultimately producing 2 runs on GPT-4 version 0613 (12/5/23 and 12/21/23), 5 runs on GPT-4v version 1015 (12/18/23 and 4 runs on 12/21/23) and 22 runs on GTP-4-Turbo

version 1106-Preview (12/21/23, 1/18/24, and 5 runs each on 1/22/24, 1/30/24, 1/31/24, and 2/8/24). The same test set, prompts, and python program were used with one exception: before the runs on 1/18/24 were completed we updated the OpenAI Python API library from version 0.27.9 to version 1.7.2[10]. This allowed us to specify a consistent seed parameter for GPT-4-Turbo's random number generator -- we used 42 -- to reduce nondeterminism. After the library upgrade, we were able to track the GPT-4 Turbo's system fingerprint for each run which is intended to represent the overall environmental configuration thereby enabling identification of environmental changes. All fingerprints were identical.

### Results

Figures 3 and 4 show performance over time with the vertical dotted line representing the API upgrade. The error bars on days where multiple runs for a single model occurred can be viewed as a measure of nondeterminism, which consistently appears across all performance metrics, even after using a consistent seed parameter. The yellow ribbon indicates the standard deviation of the 20 GPT-4-Turbo runs occurring on 1/22/24 – 2/8/24. Significant prompt 1 NPV and sensitivity drift appears to be visible before 1/22/24 for GPT-4-Turbo and GPT-4v. Prompt 1 PPV performance has substantially less drift. Drift appears relatively high in all Prompt 2 metrics, except for Benign classification NPV. It is possible that nondeterminism could explain the drift. Drift appears to significantly reduce after 1/22/24.

Table 1 shows the confusion matrix for the last 20 runs of the test set after 1/18/24 together with the standard deviation for each metric calculated using each individual run's metric as a datapoint. Given that these final 20 runs have more consistent performance characteristics, these standard deviations represent the nondeterminism of GTP-4.

### Discussion

We initially assumed that one run of our test dataset would be sufficient to establish its performance characteristics, and that drift and nondeterminism were not important to analyze. However, as our results show, even when tuning model parameters to minimize nondeterminism (e.g., setting temperature to 0 and setting a consistent seed), there is still substantial nondeterminism that must be considered when assessing performance. Doing so requires repeatedly running test datasets to assess metric by metric variability. This variability must be considered when using a metric to determine the suitability for introducing functionality to a care delivery process. If this is not done, there is a risk of overweighting a favorable result that may have occurred by random chance.

The risk of drift also appears to be an important factor to consider. In our tests, considerable drift appeared to occur before 1/22/24. After this date, performance appeared to stabilize with intraday differences explainable through nondeterminism. However, there is no guarantee that drift will not reoccur. This suggests the need to monitor deployed model performance over time. Hazards that could be introduced if a model drifts unexpectedly should also be considered and mitigated before LLM deployment into clinical workflows.

Once drift stabilized after 1/22/24, we demonstrated the ability for programmatic GPT-4 based chain-of-thought prompts to detect the presence or absence of a variant-specific functional assay within a scientific publication with high sensitivity and PPV and low nondeterminism. This is the first step in increasing the efficiency of variant review for functional studies, as literature searches using standard methodologies can return literature where the variant of interest is not discussed or does not have informative data. This capability could assist in prioritizing papers likely to have relevant evidence. The GPT-4 prompts were also able to detect whether an article had clear evidence supporting a benign or pathogenic assessment for a specific variant with high mean NPV and low NPV nondeterminism. These capabilities could help deprioritize papers unlikely to have functional evidence that impacts the classification of a variant. PPV for identifying evidence in papers supporting a pathogenic or benign call where too low to consider using this prompt sequence for fully automated assessment in this workflow. There are numerous potential improvements that could be explored which might raise performance to this level.

One limitation of the approach is the current inability to identify functional data present solely in figures or supplemental material. Expanding the python program and prompt sequences to consider these data elements will improve the utility of the tool, especially as high-scale functional assays typically do not contain the functional evidence solely in tables or text from the main part of the manuscript. It is possible that supplying figures to newer models capable of accepting them would improve performance.

Continued iteration of prompting strategies, including incorporating gene-specific logic into prompts and adding multi-shot examples to prompts could also significantly improve the ability to classify and extract evidence from articles, thus improving this technique's power. Incorporating explanations of mistakes GPT made processing our test set (few shot learning) may also improve performance. It is possible that domain specific models could improve results.

Functional evidence is also only one of several types of evidence that variant classification guidelines call for extracting from articles. Capabilities to extract other forms of evidence, especially proband and family segregation data, will be needed to comprehensively assess and prioritize articles for review while identifying relevant information.

## Conclusion
Our data provides support for the contention that GPT-4 Turbo can be used to improve clinical workflows but does not support its use in fully automated variant assessment. It indicates that risks associated with drift and nondeterminism must be considered in the clinical context, specifically measuring nondeterminism before implementation and monitoring drift once implemented.

## Data Availability
Data and results are available here: https://github.com/mgbpm/ai-genetics-functional-paper-supplements/releases/tag/AIGeneticsFunctional_Supplements_1.0.1

## Acknowledgements
We would like to thank Elizabeth W. Karlson, MD, Adam B. Landman, MD, Nallan Sriraman, Matthew Butler, Jonathan Hamill, Fei Wang, Eugene Mysen, Emma Perez, Anna Nagy, Benjamin M Scirica, Alexander J Blood, MD, Ozan Unlu, MD, William Gordon, MD, Jane Yung-jen Hsu, Ming-Yi Chin, and Yang-Sheng Lin for their insights, support and contributions to this effort. We thank Microsoft for supplying credits for GPT4 series model usage that were used in this study.

## Conflict of Interests
Samuel Aronson, Jiyeon Shin, Charlotte Mailly, and Michael Oates research grants and similar funding via Brigham and Women's Hospital from Better Therapeutics, Boehringer Ingelheim, Eli Lilly, Milestone Pharmaceuticals, NovoNordisk and PICORI. Samuel Aronson, Michael Oates, Kalotina Machini, Emma Henricks, and Matthew Lebo reporting NIH funding through Mass General Brigham. Mr. Aronson reports serving as a paid consultant for Nest Genomics.

## References
1.  Brnich, S. E. *et al.* Recommendations for application of the functional evidence PS3/BS3 criterion using the ACMG/AMP sequence variant interpretation framework. *Genome Med.* 12, 3 (2019).

2. Brnich, S. E., Rivera-Muñoz, E. A. & Berg, J. S. Quantifying the potential of functional evidence to reclassify variants of uncertain significance in the categorical and Bayesian interpretation frameworks. *Hum. Mutat.* 39, 1531–1541 (2018).
3. Kanavy, D. M. *et al.* Comparative analysis of functional assay evidence use by ClinGen Variant Curation Expert Panels. *Genome Med.* 11, 77 (2019).
4. Vanoye, C. G. *et al.* High-Throughput Functional Evaluation of KCNQ1 Decrypts Variants of Unknown Significance. *Circ Genom Precis Med* 11, e002345 (2018).
5. Gasperini, M., Starita, L. & Shendure, J. The power of multiplexed functional analysis of genetic variants. *Nat. Protoc.* 11, 1782–1787 (2016).
6. OpenAI. GPT-4 Technical Report. *arXiv [cs.CL]* (2023).
7. Bang, Y. *et al.* A Multitask, Multilingual, Multimodal Evaluation of ChatGPT on Reasoning, Hallucination, and Interactivity. *arXiv [cs.CL]* (2023).
8. Ouyang, S., Zhang, J. M., Harman, M. & Wang, M. LLM is Like a Box of Chocolates: the Non-determinism of ChatGPT in Code Generation. *arXiv [cs.SE]* (2023).
9. Freeman, P. J., Hart, R. K., Gretton, L. J., Brookes, A. J. & Dalgleish, R. VariantValidator: Accurate validation, mapping, and formatting of sequence variation descriptions. *Hum. Mutat.* 39, 61–68 (2018).
10. *openai-python: The official Python library for the OpenAI API*. (Github).

Table 1. Performance of GTP-4 Turbo assessment of the presence of functional data and the classification of the functional evidence when present.

| | | Manual classification | | | | | Manual classification | | | | |
|---|---|---|---|---|---|---|---|---|---|---|---|
| | | Functional Evidence Present | No Assays Detected | | | | Pathogenic | Intermediate/ Inconclusive | Benign | PPV | NPV |
| GPT-4 Prompts | Functional Evidence Present | 867 | 40 | PPV: 95.6% (2.0%) | | GPT-4 Prompts | Pathogenic | 233 | 82 | 0 | 74.0% (5.0%) | 95.3% (2.8%) |
| | No Assays Detected | 73 | 460 | NPV: 86.3% (3.1%) | | | Intermediate/ Inconclusive | 26 | 304 | 21 | 86.6% (5.4%) | 75.0% (3.2%) |
| | | | | | | | Benign | 0 | 47 | 154 | 76.6% (8.6%) | 96.9% (1.2%) |
| | | Sensitivity: 92.2% (2.0%) | Specificity: 92.0% (3.9%) | | | | Sensitivity | 90.0% (6.1%) | 70.2% (4.7%) | 88.0% (4.4%) | | |

Performance calculated on the last 20 GTP-4 Turbo runs (5 per day for 4 days). Failure to generate a result from GTP-4 Turbo are classified as "No Assay Detected". Mapping of table labels to GTP-4 Turbo output: Pathogenic = "Pathogenic Evidence"; Intermediate = "Evidence of Intermediate Function"; Inconclusive = "Results are Inconclusive"; Benign = "Benign Evidence"; PPV=positive predictive value; NPV = negative predictive value. Standard deviations are calculated based upon the metrics produced in the 20 runs.

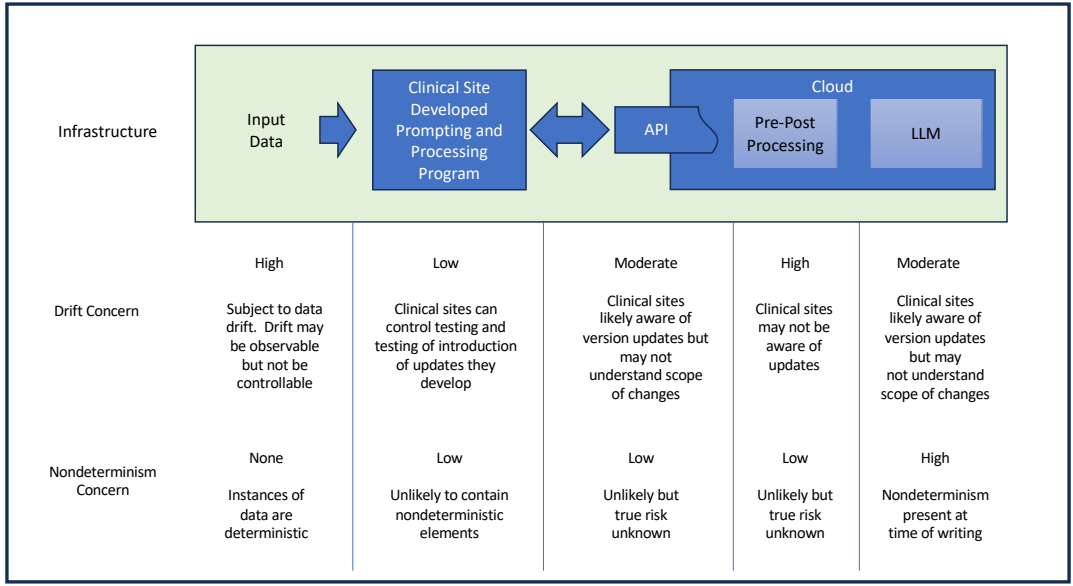

Figure 1: Infrastructure components subjected to drift and nondeterminism

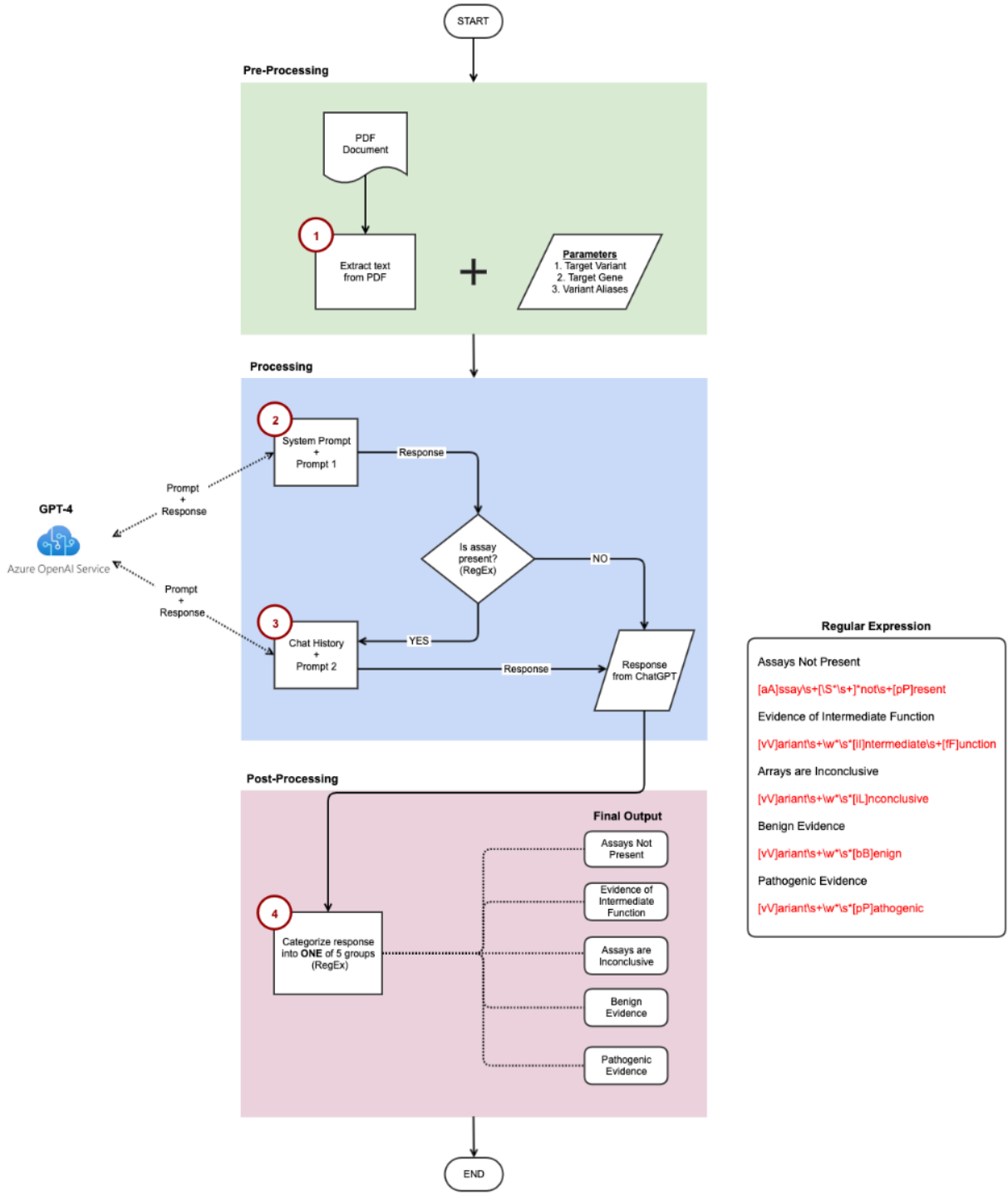

Figure 2. Article processing workflow. Articles are processed as follows: 1) extract text from article; 2) GPT-4 prompt for identifying presence of functional evidence; 3) GTP-4 prompt for classification of the variant-level functional data; 4) post-processing of responses from GTP-4.

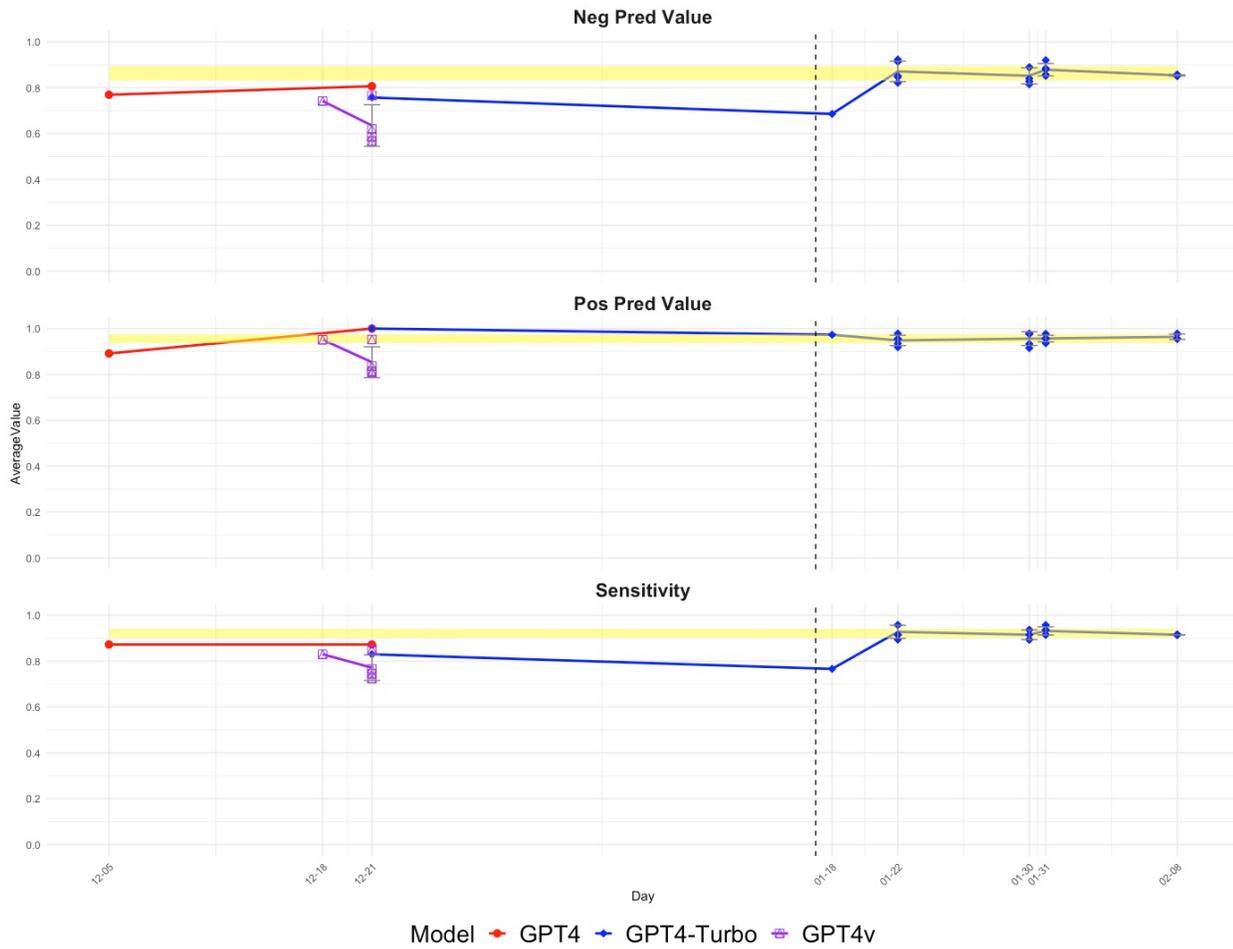

Figure 3. First prompt performance identifying articles containing functional data for variant of interest. Yellow ribbons represent the standard deviation of data from the last 20 GTP-4 Turbo analyses. Dotted lines indicated the timing of the PI upgrade and implementing setting a seed.

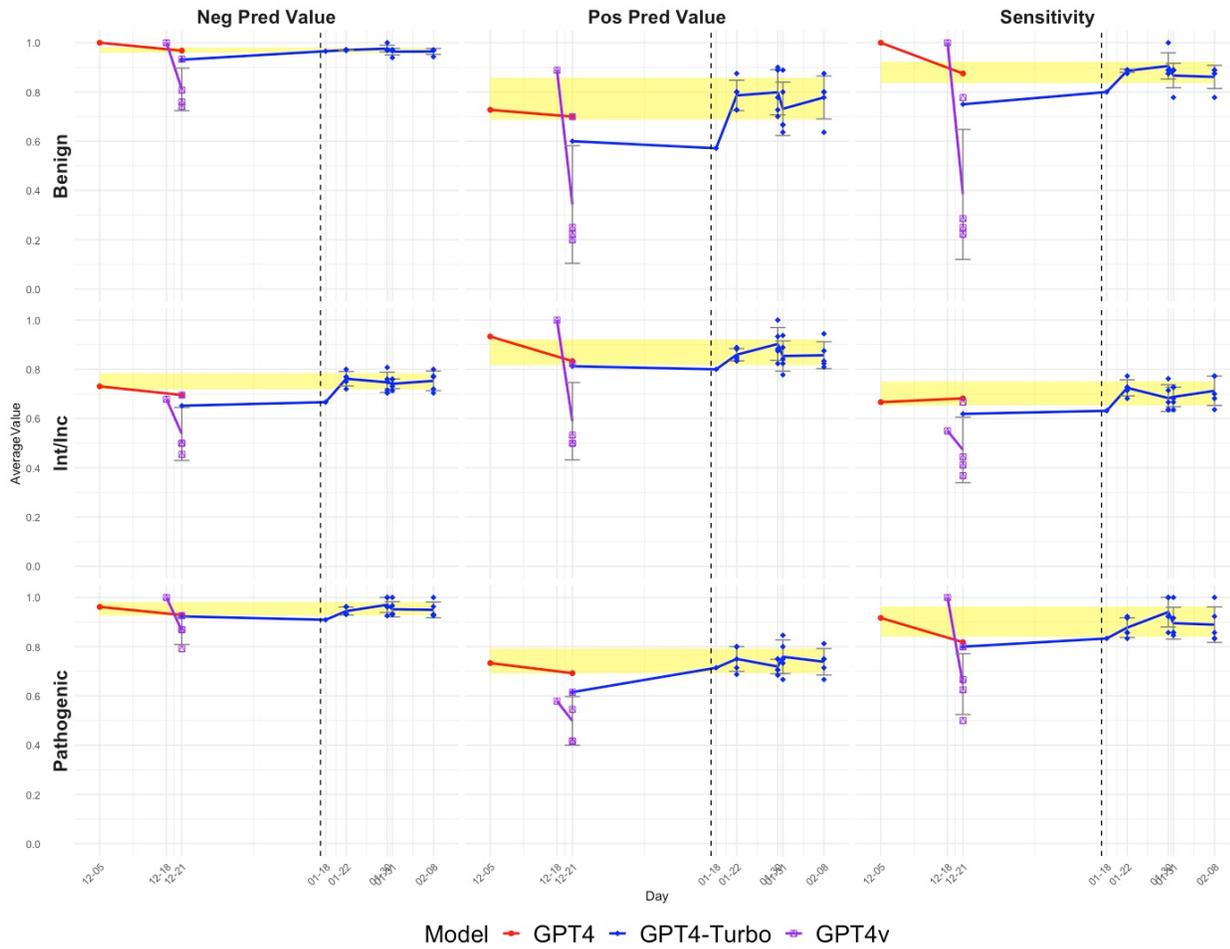

Figure 4. Performance of second prompt classifying functional evidence for variants for which functional data was identified in first prompt. Yellow ribbons represent the standard deviation of data from the last 20 GTP-4 Turbo analyses. Dotted lines indicated the timing of the PI upgrade and implementing setting a seed.

Supplemental Table 1: Prompts and processing logic.

| Parameter | Description |
|---|---|
| $param_variant | A target variant per article |
| $param_gene | A target gene per article |
| $param_variant_aliases | A list of nomenclature aliases equivalent to the target variant |
| $content | Content of article in plain text |

| Step | Prompt |
|---|---|
| System Prompt | "You are a computer program working for a geneticist. You will analyze the genetic variant "$param_variant" in the "$param_gene" gene in the context of patient care by researching academic publications. The genetic variant, "$param_variant", is also known as "$param_variant_aliases". You will evaluate the variant's pathogenicity in publications by searching for evidence of functional assays (as defined by animal studies, human studies, in vivo, in vitro, cellular, molecular or other studies for protein or enzymatic function, localization or expression, or any other test that assesses a variant's impact on gene function) for specified variants." |
| Prompt 1 | "In the publication provided below delimited by triple backticks, check if the $param_variant variant was tested using any animal studies, human studies, in vivo, in vitro, cellular, molecular or other studies for protein or enzymatic function, localization or expression, or any other test that assesses a variant's impact on gene function. If yes, output the result of these studies for the $param_variant variant. If there is no information about any of these studies about this variant, output "assay information not present". Publication: ```$content```"<br><br>The above prompt is run up to 3 times with $param_variant for the different nomenclatures below until assays are found or all fail:<br>1. The system searches for and uses the longest nomenclature used in the article from the input alias list. This could be either the target variant or one of its nomenclature aliases.<br>2. If step 1 does not return assays, the system uses the target variant input parameter.<br>3. If step 2 does not return assays, the system prefixes the Gene to the target variant input parameter. |
| Prompt 2 | "If the previous output indicates that the variant $param_variant is pathogenic (or significantly alters protein or enzymatic function, localization or expression), say "Assays Indicate Variant Is Pathogenic". If the results indicate that the variant is benign or similar to wild type (WT) or does not impact protein function, say "Assays Indicate Variant is Benign." If the results indicate the variant has partial function, say "Assays indicate Variant has Intermediate Function". If the results indicate that the assays are inconclusive, say "Assays are inconclusive"." |

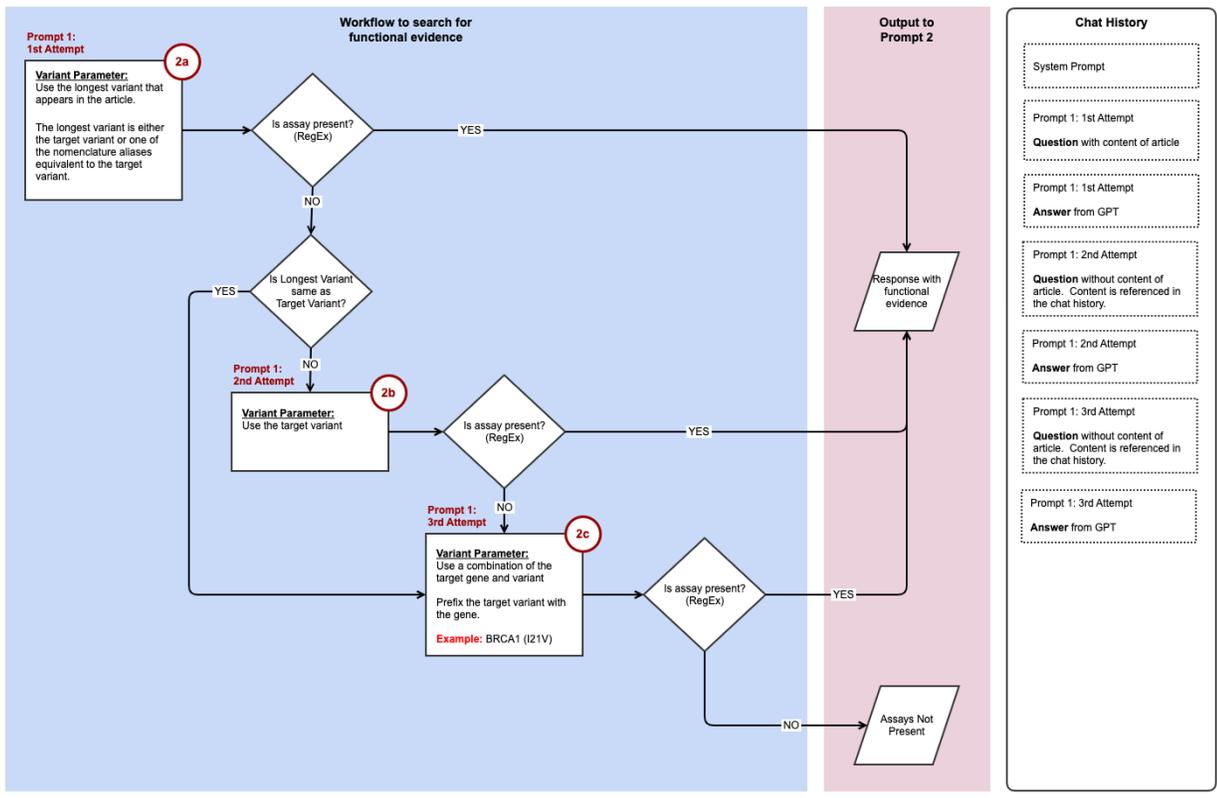

Supplemental Figure 1. Detailed Breakdown of Second Processing Step: Interactive process for determining if functional evidence is present.